
\input harvmac.tex
%
%
%
%
%
%
%
\lref \rBKMS { V. V. Bazhanov, R. M. Kashaev, V. V. Mangazeev and Yu.
G. Stroganov, {\sl $({\bf Z_N}\times)^{n-1}$ Generalization of the
Chiral Potts Model},
Commun. Math. Phys. {\bf 138} (1991) 393. }
\lref \rDJMM { E. Date, M. Jimbo, K. Miki and  T. Miwa,
{\sl New $\cR$-matrices associated with cyclic representations of
$\ \cU_q(A_2^{(2)})$}, Kyoto preprint RIMS-706 and
{\sl Generalized chiral Potts models and minimal cyclic
representations of
$\ \cU_q \left(\widehat{gl}(n,\CC )\right) $,}
Commun. Math. Phys. {\bf 137}  (1991) 133.}
\lref \rDK {  C. De Concini and V.G. Kac,
{\sl Representations of quantum groups at roots of 1,}
Progress in Math. {\bf 92} (1990) 471 (Birkh\"auser).}
\lref \rKerDipl { T. Kerler,
{\sl Darstellungen der Quantengruppen und Anwendungen,}
Diplomarbeit, ETH--Zurich, August 1989. }
\lref \rLus { G. Lusztig,
{\sl Quantum groups at roots of $1$,} Geom. Ded. {\bf 35} (1990) 89.}
\lref \rRA {  P. Roche and  D. Arnaudon,
{\sl Irreducible representations of the quantum
analogue of $SU(2)$,}
Lett. Math. Phys. {\bf 17} (1989) 295.}
\lref \rRosC { M. Rosso, {\sl Quantum groups at root of 1 and tangle
invariants,} in Topological and geometrical methods in field theory,
Turku, Finland May-June 1991, ed. by J. Mickelsson and O. Pekonen.}
\lref \rRT { N. Reshetikhin and V. Turaev, {\sl Invariant of 3
manifolds via link polynomials and quantum groups,} Invent. Math. {\bf
103} (1991) 547.}
%
%
%
%
%
%
\def\nopage0{\pageno=0 \footline={\ifnum\pageno<1 {\hfil} \else
              {\hss\tenrm\folio\hss}   \fi}}
%
%

%
%
%

%
%

          \def\cQ{{\cal Q}}          \def\cR{{\cal R}}
                    \def\cU{{\cal U}}

%
%

\def\CC{\rlap {\raise 0.4ex \hbox{$\scriptscriptstyle |$}}
\hskip -0.1em C}
\def\FF{\hbox to 8.33887pt{\rm I\hskip-1.8pt F}}
\def\NN{\hbox to 9.3111pt{\rm I\hskip-1.8pt N}}
\def\PP{\hbox to 8.61664pt{\rm I\hskip-1.8pt P}}
\def\QQ{\rlap {\raise 0.4ex \hbox{$\scriptscriptstyle |$}}
{\hskip -0.1em Q}}
\def\RR{\hbox to 9.1722pt{\rm I\hskip-1.8pt R}}
\def\ZZ{\hbox to 8.2222pt{\rm Z\hskip-4pt \rm Z}} 
\def\uq{\cU_q(sl(2))}
\def\uni{universal $\cR$-matrix}
\def\unis{universal $\cR$-matrices}
\font\large=cmr10 scaled \magstep1
\font\Large=cmr10 scaled \magstep2
\def\enslapp{{\sevenrm E}N{\large S}{\Large L}{\large A}P{\sevenrm P}}
%
\line{\hfil \enslapp-A-461/94 }
\vfill
\Title{}{\vbox{\centerline{On Automorphisms and Universal $\cR$-matrices}
      \vskip2pt\centerline{at Roots of Unity}}}

\centerline{Daniel Arnaudon\footnote{$^*$}
{arnaudon@lapphp1.in2p3.fr}}
\centerline{Laboratoire de Physique Th\'eorique \enslapp
\footnote{$^1$}{
URA 14-36 du CNRS, associ\'ee \`a l'E.N.S. de Lyon et au L.A.P.P.
d'Annecy-le-Vieux.}, }
\centerline{ Chemin de Bellevue BP 110, 74941 Annecy-le-Vieux Cedex,
France}

\vfill

Invertible universal $\cR$-matrices of quantum Lie algebras
do not exist at roots of unity.
There exist however quotients
for which intertwiners of tensor
products of representations always exist, i.e. $\cR$-matrices exist in
the representations.
One of these quotients, which is finite dimensional,
has a universal $\cR$-matrix.
In this paper, we answer the following question: on which condition
are the different quotients of \ $\uq$  (Hopf)-equivalent?
In the case when they are equivalent, the universal $\cR$-matrix of
one can be transformed into a universal $\cR$-matrix of the other.
We prove that this happens only when $q^4=1$, and we explicitly give the
expressions for the automorphisms and for the transformed universal
$\cR$-matrices in this case.
\vfill

\rightline{\enslapp-A-461/94}
\rightline{hep-th/9403110}
\rightline{March 94}

\baselineskip=17pt

\Date{}

\newsec{Introduction}
The following section is devoted to some generalities and definitions.
We recall there why an invertible \uni\  cannot exist at
roots of unity on the whole $\uq$. Then we define some quotients
$\cQ_{A,B}$  of $\uq$
on which this obstruction does not exist.
In section 3,
we prove that there exists algebra automorphisms of $\uq$ that lead to
isomorphisms between these quotients.
We also prove that
there exists a continuous group of Hopf-algebra automorphisms of the
the the central Hopf-subalgebra (called $Z_0$)
of $\uq$  that exchange the defining relation of these quotients.
Then we investigate under which conditions these
automorphisms can be extended to Hopf algebra isomorphisms of the
whole quotients $\cQ_{A,B}$.
We find that this is possible only when $q^4=1$. In section 4, we explicitly
give the isomorphisms that link the quotients in the case $q^4=1$,
and deduce from this an expression for a \uni\ on any quotient.

\newsec{Generalities}
The quantum algebra
$\uq$ is defined by the generators $k$, $k^{-1}$, $e$, $f$,
and the
relations
\eqn\eSL{
\eqalign{
& kk^{-1}=k^{-1}k=1 ,\cr
& [e,f]={k-k^{-1} \over q-q^{-1}}, \cr }
\qquad \qquad
\eqalign{
& kek^{-1}=q^2 e ,\cr
& kfk^{-1}=q^{-2}f .\cr}}

The coproduct, co-unit and antipode endow $\uq$ with a Hopf algebra
structure.
The
coproduct $\Delta$ is the morphism of algebras
$ \uq \longrightarrow \uq\otimes\uq$
defined by
\eqn\eDelta {
\eqalign {
& \Delta(k)=k\otimes k \cr
& \Delta(e)=e\otimes 1      + k\otimes e \cr
& \Delta(f)=f\otimes k^{-1} + 1\otimes f \;,\cr  }}
while the opposite coproduct $\Delta '$ is $\Delta '= P \Delta$, where $P$
is the permutation map $P x\otimes y =y \otimes x$.

The co-unit $\epsilon$ is the morphism of algebras $\uq \longrightarrow \CC$
defined by
\eqn\counit{\epsilon(k)=1, \qquad \epsilon(e)=\epsilon(f)=0}
and the antipode $S$ is the antimorphism of algebras
$\uq\longrightarrow\uq$ given by
\eqn\eantipode{S(k)=k^{-1}, \qquad S(e)=-k^{-1}e, \qquad S(f)=-fk.}
\par From now on,
$q$ will be a primitive $m'$ root of unity ($q^2\neq 1$), and $m$ will
be set to $m'$ if $m'$ is odd, or to $m'/2$ if $m'$ is even. From the
relations \eSL\ we deduce that, besides the deformation $C$ of
the quadratic Casimir, the operators $x\equiv e^m$, $y\equiv f^m$ and
$z^{\pm 1}\equiv k^{\pm m}$
belong to the centre $Z$ of $\uq$ \rRA. These operators and $C$ actually
generate this centre; they altogether satisfy a polynomial relation
of degree $m$ in $C$, and $1$ in  $x$, $y$ and $z^{\pm 1}$ \rKerDipl.
The (Hopf) algebra generated by $x$, $y$ and $z^{\pm 1}$ will be
denoted by $Z_0$ \rDK.

Let us now recall why a \uni\ cannot exist when $q$ is a root of
unity \rDJMM. A \uni\ is an {\it invertible}
element $\cR$ of $\uq\otimes \uq$ such that
\eqn\euni{\cR \Delta(a)=\Delta '(a)\cR \qquad \forall a\in\uq.}
On the tensor product of two irreducible representations $\rho_1$ and
$\rho_2$, one would have
\eqn\eunir{(\rho_1\otimes\rho_2)(\cR) (\rho_1\otimes\rho_2)(\Delta(a))=
(\rho_1\otimes\rho_2)(\Delta '(a))(\rho_1\otimes\rho_2)(\cR)
\qquad \forall a\in\uq,}
with $\rho_1\otimes\rho_2(\cR)$ invertible. It follows from \eSL\ and
\eDelta\ that
\eqn\edeltax{
\eqalign{\Delta(x) &= x\otimes 1 + z\otimes x ,\cr
         \Delta '(x) &= x\otimes z + 1\otimes x ,\cr}
\qquad
\eqalign{\Delta(y) &= y\otimes z^{-1} + 1\otimes y ,\cr
         \Delta '(y) &= y\otimes 1 + z^{-1}\otimes y .\cr}}
Since $x$, $y$ and $z$ are central, they act by scalar on irreducible
representations $\rho_1$, $\rho_2$,
and hence so do $\Delta(x)$, $\Delta(y)$ and $\Delta(z)$ on
$\rho_1\otimes\rho_2$.
Taking $a=x$ (resp. $a=y$) in \eunir, and simplifying by
$\rho_1\otimes\rho_2(\cR)$, we then get
\eqn\econtradiction{\eqalign{
\rho_1(x)+\rho_1(z)\rho_2(x) & = \rho_2(x) +\rho_2(z)\rho_1(x) \cr
\rho_1(y)\rho_2(z^{-1})+\rho_2(y) & = \rho_2(y) \rho_1(z^{-1})+\rho_1(y) \cr
}}
But $x$, $y$ and $z$ can take arbitrary complex values on irreducible
representations \rRA, which is in contradiction with \econtradiction.
So there exist no \uni\ on $\uq$ when $q$ is a root of unity.

We can however define quotients of $\uq$ on which this contradiction
does not exist.

The simplest is the so-called finite dimensional quotient defined by ($x=0$,
$y=0$, $z=\pm 1$) \rLus. On this quotient, a \uni\ can be defined
\rRosC, \rRT.

The condition $z=\pm 1$ can be relaxed. We denote this new quotient by
$\cQ_{0,0}$. An almost \uni\ can be defined on the
quotient $\uq /(x=y=0)$ \rRosC.
The irreducible representations of maximal dimension $m$
of this quotient are parametrized
by the value of $z$.

More general quotients with no obstruction are given by
\eqn\equot{\cQ_{A,B}  =  \uq /(x=A(z-1), y=B(1-z^{-1}))}
for complex values of $A$ and $B$.
On quotients of this family, the relations \econtradiction\ are satisfied.
The previously considered  quotient belongs to this family since it
corresponds to $A=B=0$.
It is not known whether a \uni\ can be defined on this type of
quotient. But it is well-known that $\cR$-matrices exists for
representations of these quotient: for each pair of irreducible
representations $\rho_1$ and $\rho_2$, there exist
$\cR_{\rho_1,\rho_2}$ such that
\eqn\eronrep{\cR_{\rho_1,\rho_2} (\rho_1\otimes\rho_2)(\Delta(a))=
(\rho_1\otimes\rho_2)(\Delta '(a))\cR_{\rho_1,\rho_2}
\qquad \forall a\in\uq.}
The parameters of the representations, considered as representations
of $\uq$ indeed
satisfy an algebraic equation  (the equation of the spectral variety of
the related physical model)
corresponding to the existence of
an intertwiner \rDJMM, \rBKMS.
This equation is equivalent to
the relations defining the
quotient.

Finally we can consider the quotient $\uq /(z-1)$. The structure of
this quotient is different from the structure of the quotients \equot.
Irreducible representations of maximal dimension $m$
of this quotient are parametrised by two
continuous parameters (the values of $x$ and $y$) instead of one.
This is the largest quotient of $\uq$ for which the fusion rules of
irreducible representations are commutative.

The ideals generated by the relations that define these quotients are
also co-ideals, so that these quotients are all Hopf algebras.

\newsec{On automorphisms of $\uq$}
In this section, we look for Hopf algebra automorphisms of $\uq$ that
would send one of the quotients to another.
Our motivation is the following: as we will see later,
there exist Hopf algebra automorphisms of $Z_0$, the central
Hopf-subalgebra of
$\uq$ generated by $x$, $y$ and $z^{\pm 1}$, that send the relations
defining one quotient \equot\ to the relations defining another
quotient, with different values of $A$ and $B$.
On the other hand, we can also find algebra
automorphisms of $\uq$ that send a quotient of type \equot\  to another
quotient of the same type.

\subsec{Existence of automorphisms of $\uq$ linking the different quotients}
Inspired by the derivations defined in \rDK, we construct the
following derivation on $\uq$ (in fact on the  extension of $\uq$
involving the rational functions in $z$):
\eqn\ederiv{\eqalign{
dk & = - k(bx+ay) \;, \cr
de & = a (q-q^{-1})^{-2} f^{m-1} (qk-q^{-1}k^{-1}) \;, \cr
df & = b (q-q^{-1})^{-2} e^{m-1} (q^{-1}k-qk^{-1}) \;, \cr}}
with
\eqn\ederivwith{\eqalign{
a & = D^{-1} z \left( a'(\lambda(z+1)+AB) - b' A^2z \right) \;,\cr
b & = D^{-1}   \left( - a'B^2 + b' (\lambda(z+1) + ABz) \right) \;,\cr
\lambda &= q^m (q-q^{-1})^{-2m} \;,\cr
D & = \lambda (1+z)^2 (\lambda + AB) \;.\cr
}}
On the centre of $\uq$, the action of this derivation is given by
\eqn\ederivcentre{\eqalign{
dC & = 0 \;,\cr
dz & = - mz (bx+ay) \;, \cr
dx & = m \lambda a (z-z^{-1}) \;, \cr
dy & = m \lambda b (z-z^{-1}) \;. \cr
}}

Following the flows given by these derivations, we get automorphisms
of $\uq$.

On the family of quotients $\cQ_{A,B}$, these derivations provide a
parallel transport of the algebra structure, the action on the
parameters $A,B$ being $dA=ma'$, $dB=mb'$.
Following the flow given by $d$, we then get isomorphisms between the
quotients $\cQ_{A,B}$ (as algebras).

\subsec{Automorphisms of $Z_0$  (as Hopf algebra)}
As explained before as a motivation,
it is easy to find automorphisms of the central part $Z_0$ of $\uq$,
whose action is to send the defining relation of a quotient \equot\
to another. These
automorphisms are moreover Hopf algebra automorphisms. Such an
automorphism $\phi$ is given by
\eqn\ephi{\eqalign{
\phi(z) & = z\;, \cr
\phi(x) & = a_1 x + a_2 yz + a(z-1) \;, \cr
\phi(y) & = a_3 xz^{-1} + a_4 y + b(1-z^{-1}) \;,\cr
}}
with $a_i$, $a$ and $b$ in $\CC$.

\subsec{Existence of (Hopf algebra) automorphisms of $\uq$}
The existence of automorphisms of $\uq$ linking the different
quotients \equot, on one hand, and, on the other hand, the existence of
Hopf algebra automorphisms of $Z_0$ sending the relations defining
these quotients one to another raise the question of the existence of
Hopf algebra automorphisms of the whole $\uq$, that would relate the
different quotients. Or do isomorphisms of Hopf algebra link the
different quotients?
We will prove here that such Hopf algebra isomorphisms exist only
when $q^4=1$.

Since the family of quotients \equot\ is parametrized by two continuous
parameters, it is natural to look for a Lie group of
automorphisms of $\uq$. If such a group of
automorphisms exists, its Lie algebra provides derivations on $\uq$.
If these automorphisms respect the Hopf algebra structure, then so do
the derivations.

Let us look for a derivation $d$ of $\uq$ such that
\eqn\ehopfder{\eqalign{
\Delta(dk) & = dk\otimes k + k\otimes dk \;,\cr
\Delta(de) & = de\otimes 1 + dk\otimes e + k\otimes de \;,\cr
\Delta(df) & = df\otimes k^{-1} + f\otimes d(k^{-1})+ 1\otimes df \;,\cr
}}

Since automorphisms preserve the centre, we first look for derivations
that preserve the centre . We write  the most general
\eqn\edz{dz=\sum_{p_1,p_2 \geq 0} \sum_{p_3} \sum_{p_4=0}^{m-1}
a_{p_1,p_2,p_3,p_4}
x^{p_1} y^{p_2} z^{p_3} C^{p_4} \;, }
and demand
\eqn\edzhopf{\Delta(dz)  = dz\otimes z + z\otimes dz \;. }
(In \edz, the degree of the quadratic Casimir $C$ is
limited by the polynomial relation it satisfies with $x$, $y$ and $z$. )

\medskip \noindent {\bf Proposition: } {\sl If $d$ is a derivation of
Hopf algebra, then $dz = 0$. }

\noindent {\it Proof:}
The presence of powers of $C$ in \edz\ would lead to terms in $e$ and
$f$ in both sides of the tensor products in $\Delta(dz)$. By recursion
in decreasing $p_4$, we then get that the coefficient is $0$ unless
$p_4=0$. A similar argument leads to $p_1+p_2\leq 1$. We finally get
$dz=0$ by inspection.
\medskip

For $dx$ and $dy$, we write a formula similar to \edz. We impose
\eqn\edxyhopf{\eqalign{
\Delta(dx) & = dx\otimes 1 + dz\otimes x + z\otimes dx \;,\cr
\Delta(dy) & = dy\otimes z^{-1} + y\otimes d(z^{-1})+ 1\otimes dy \;.\cr
}}

\medskip \noindent {\bf Proposition: } {\sl If $d$ is a derivation of
Hopf algebra, then its action on $x$ and $y$ is
\eqn\edxy{\eqalign{
dx & = a_1 x + a_2 yz + a(z-1) \;, \cr
dy & = a_3 xz^{-1} + a_4 y + b(1-z^{-1}) \;.\cr}}
}

\noindent {\it Proof:}
By the same kind of arguments as before.
\medskip

Thus, derivations of the Hopf algebra $\uq$, if they exist, leave
$Z_0$ invariant. They lead to the automorphisms \ephi\ of the Hopf
algebra $Z_0$. We now need to extend the action of these derivations
to $\uq$. Since $dz=0$, and hence $dk=0$,
we already know that we are not going to
obtain the derivations \ederiv\ as derivations of Hopf algebra.

We now consider
\eqn\ede{de=\sum_{p_1,p_2 \geq 0} \sum_{p_3}
a_{p_1,p_2,p_3}
e^{p_1} f^{p_2} k^{p_3}  }
and a similar formula for $df$,
and we impose \ehopfder.

\medskip \noindent {\bf Proposition: }{\sl

\item{\bf a.} {If $m\neq 2$,
the only derivations of Hopf algebra are the trivial derivations
$de=ae$, $df=-af$ and $dk=0$.
They lead to automorphisms that are just
rescalings of $e$ and $f$
leaving the product $ef$ invariant. These automorphisms of $\uq$
provide  isomorphisms between all the quotients $\cQ_{A,B}$ that have
the same value for the product $AB$.}

\item{\bf b.} {In the case $m=2$, there exists a
group of  Hopf algebra isomorphisms linking the different quotients.
Explicit
expressions are given in the following section. }

}

\noindent {\it Proof:}
Reasoning as for $dx$ and $dy$, we get formulas similar to \edxy\ for
$de$ and $df$,
replacing $x$, $y$ and $z$ by $e$, $f$ and $k$ respectively. But these
elements are now non central and the derivation has to be compatible
with the relations \eSL. In particular,
\eqn\ekdek{k\ de\ k^{-1} = q^2 \ de \;,
\qquad k\ df\ k^{-1} = q^{-2} \ df \;,}
so that the terms in $1$ and $k$ are eliminated, whereas the term in
$f$ in $de$ (resp. the term in $e$ in $df$) can be kept only if
$q^2=q^{-2}$.
In this case, there is still no non trivial
derivation of the whole Hopf algebra
$\uq$ (i.e. with respect to both \eSL\ and \ehopfder).   The problem
arises with $d([e,f])$ which involves $e^2$ and $f^2$. Nevertheless,
if we use the relations $e^2=A(z-1)$ and $f^2=B(1-z^{-1})$, we get
$d([e,f])=0$, compatible with $dk=0$. We then have $dA\neq 0$ and
$dB\neq 0$. So $d$ can be seen as a connection lifting $dA$ and $dB$
(the basis being the space of parameters $A,B$) to the bundle defined
by the family of quotients $\cQ_{A,B}$. Using $d$, or by direct
calculation, we get isomorphims of Hopf algebras between the
quotients.

\medskip
{\it Remark:} As suggested by M. Bauer,
there exist derivations of $\uq$ that do not respect the
grading \ekdek. These derivation are of course such that $dk\neq 0$.
Hence they do not satisfy \ehopfder.

\newsec{Isomorphisms and \unis\ of quotients of $\uq$ at $q^4=1$}
The following $\phi$ defines  Hopf algebra isomorphisms
from the quotient $\cQ_{A,B}$ of $\uq$
with $q^4=1$ to the quotient $\cQ_{A',B'}$
\eqn\ephihopf{\eqalign{
k_{A',B'} &= \phi(k_{A,B}) = k_{A,B} \;,\cr
e_{A',B'} &= \phi(e_{A,B}) = N^{-1} (e_{A,B}+\alpha f_{A,B} k_{A,B}) \;,\cr
f_{A',B'} &= \phi(f_{A,B}) = N^{-1} (f_{A,B}+
                                     \beta e_{A,B} k_{A,B}^{-1}) \;, \cr
N       & = (1+\alpha\beta+4q(\beta A-\alpha B))^{1/2} \cr
}}
with
\eqn\ephihopfAB{\eqalign{
A' &= N^{-2} (A -{1\over 2}\alpha q -\alpha^2 B) \;, \cr
B' &= N^{-2} (B +{1\over 2}\beta q -\beta^2 A)  \;. \cr
}}

For clarity, we put indices indicating which quotient the operators
belong to.

We will
use these isomorphims to define an $\cR$-matrix on the quotients
$\cQ_{A,B}$. The $\cR$-matrix on the quotient $\cQ_{0,0}$ is
\eqn\eRzero{\cR_{0,0} = (1\otimes 1 -2q\ e_{0,0}\otimes f_{0,0})\
                    q^{h\otimes h \over 2}  \;,}
where we have introduced the new generator $h$ in $\cQ_{0,0}$, such
that $k_{0,0}=q^h$, $[h,e_{0,0}] = 2e_{0,0}$ and
$[h,f_{0,0}]=-2f_{0,0}$.
Note that this is consistent with
the fact that $x_{0,0}$
and $y_{0,0}$ remain central, although $[h,x_{0,0}]=4x_{0,0}$ and
$[h,y_{0,0}]=-4y_{0,0}$, simply
because $x_{0,0}=y_{0,0}=0$ on this quotient.

Now the same $\cR$-matrix works on $\cQ_{A,B}=\phi(\cQ_{0,0})$, since
$\phi$ is an isomorphism of Hopf algebra:
\eqn\eRDphi{\eqalign{
\cR_{0,0} (\phi\otimes\phi)(\Delta(a)) & = \cR_{0,0} \Delta(\phi(a)) \cr
                                   & = \Delta' (\phi(a)) \cR_{0,0} \cr
                         & = (\phi\otimes\phi)(\Delta'(a)) \cR_{0,0} \cr
}\qquad \forall a\in\uq }
but it is not expressed in terms of the generators of $\cQ_{A,B}$.
So we write
\eqn\ephiinv{\eqalign{
e_{0,0} &= (1+\alpha\beta)^{-1/2}(e_{A,B}-\alpha f_{A,B}k_{A,B}) \cr
f_{0,0} &= (1+\alpha\beta)^{-1/2}(f_{A,B}-\beta e_{A,B}k_{A,B}^{-1}) \cr
}}
and hence the $\cR$-matrix of $\cQ_{A,B}$ is
\eqn\eRAB{\cR_{A,B} = (1\otimes 1 -{2q\over 1+\alpha\beta}
      (e_{A,B}-\alpha f_{A,B}k_{A,B}) \otimes
          (f_{A,B}-\beta e_{A,B}k_{A,B}^{-1}) )\
                    q^{h\otimes h \over 2}  \;,}
with $h$ such that
\eqn\eh{\eqalign{
k_{A,B} & = q^h \;,\cr
[h, e_{A,B}] &= 2 e_{A,B} - {4 \alpha \over 1+ \alpha \beta}
                        f_{A,B}k_{A,B} \;,\cr
[h, f_{A,B}] &= -2 f_{A,B} + {4 \beta \over 1+ \alpha \beta}
                        e_{A,B}k_{A,B}^{-1} \;.\cr
}}
(these relations are compatible with the fact that
$e_{A,B}^2$ and $f_{A,B}^2$ are central
without vanishing.)

{\it Remark 1:} Note that the $\cR$-matrix we get is not built, like
$\cR$-matrices
arising from quantum doubles, as a sum of tensor products with only $e$'s on
the left and $f$'s on the right.

{\it Remark 2:} One might expect that the knowledge of isomorphisms of
algebras only (not Hopf)
between the quotients would be enough to carry the
$\cR$-matrix from one to the other, just applying $\phi\otimes\phi$ to
both sides of \euni. The problem is that the $\cR$-matrix we start
from is not exactly a \uni\ of $\cQ_{0,0}$, since it uses the operator
$h$ that does not really belong to $\cQ_{0,0}$
(or similar tricks that use explicitly the fact that the raising and
lowering operators $e$ and $f$ are nilpotent on this quotient).
 $h$ cannot be transformed by $\phi$. The $h$ used in \eRAB\ and \eh\
is not transformed by $\phi$.

\medskip
Let us finally give the \uni\ for the quotient of $\uq$  by the
relation $z=1$, which
is not isomorphic to the others:
\eqn\eRzone{\cR_{z=1} = 1\otimes 1 + 1\otimes k + k\otimes 1 -
k\otimes k . }
This $\cR$-matrix, which is nothing but the $k$ part of the
$\cR$-matrices of \rRosC\  and \rRT,  is really a \uni.
It endows this quotient with a
quasi-triangular Hopf algebra structure.

\bigskip
{\bf Acknowledgements: }

\noindent
Michel Bauer and Frank Thuillier are
warmfully thanked for discussions and fruitful
remarks.
The author also thanks N. Reshetikhin for an interesting discussion on
automorphisms of $\uq$.

\listrefs
\bye